# First Test Results of the Trans-Impedance Amplifier Stage of the Ultra-fast HPSoC ASIC

C. Chock[a], K. Flood[a], L. Macchiarulo[a], F. Martinez-Mckinney[b], A. Martinez-Rojas[b], S. Mazza[b], I. Mostafanezhad[a], M. Nizam[b], J. Ott[b], R. Perron[a], E. Ryan[b], H. F.-W. Sadrozinski[b,1], B. Schumm[b], A. Seiden[b], K. Shin[b], M. Tarka[b], D. Uehara[a], M. Wilder[b], Y. Zhao[b]

[a] *Nalu Scientific, LLC,*
   *2800 Woodlawn Dr. Ste #240, Honolulu, HI 96822 USA*

[b] *Santa Cruz Institute for Particle Physics (SCIPP)*
   *Santa Cruz, CA 95064, USA*
   E-mail: hartmut@UCSC.edu

ABSTRACT: We present the first results from the HPSoC ASIC designed for readout of Ultra-fast Silicon Detectors. The 4-channel ASIC manufactured in 65 nm CMOS by TSMC has been optimized for 50 um thick AC-LGAD. The evaluation of the analog front end with β-particles impinging on 3x3 AC-LGAD arrays (500 um pitch, 200x200 $um^2$ metal) confirms a fast output rise time of 600 ps and good timing performance with a jitter of 45 ps. Further calibration experiments and TCT laser studies indicate some gain limitations that are being investigated and are driving the design of the second-generation pre-amplification stages to reach a jitter of 15 ps.

---

[1] Corresponding author.

**Contents**

## 1. Introduction

Low Gain Avalanche Detectors (LGADs) are thin silicon detectors (ranging from 20 to 50 μm in thickness) with moderate internal signal amplification (up to a gain of ~50) [1,2], providing timing measurements for minimum-ionizing particles with resolution of about 30 ps [3,4]. In addition, the fast rise time (as low as 180 ps for 20 μm thickness) and short full charge collection time (around 1 ns) of LGADs are suitable for high repetition rate measurements in photon science and other fields [5,6]. The first implementation will be the High-Granularity Timing Detector (HGTD) in ATLAS [7] and the Endcap Timing Layer (ETL) in CMS [8] for the LHC upgrade.

Future applications in the ePIC [9] and PIONEER [10] experiments will require high precision in both time and space. High precision spatial resolution while maintaining a 100% fill factor is provided by the increased segmentation of AC-LGADs [11] (also named Resistive Silicon Detectors RSD). This is achieved by employing an un-segmented (p-type) gain layer and (n-type) N-layer, and a di-electric layer separating the metal readout pads. The design allows to decrease the number of readout channels and the value of the sensor capacitance. Versions of AC-LGADs have shown to provide spatial resolution on the scale of few to 10's of micrometers [12].

In this paper we describe the development and first measurements of the ASIC "HPSoC" prototype optimized for fast read out of AC-LGAD. We first describe the properties of the sensor signals and derive the requirements for the readout ASICs. Then we give an overview of the basic function of the ASIC, followed by results from the electric characterization using fast calibration signals. This is followed by the investigation of the response to IR laser TCT and β particles impinging on an AC-LGAD array wire-bonded to the ASIC, and by the interpretation of the results in terms of the expected jitter.

## 2. Concept and Design of HPSoC ASIC

### 2.1 LGAD Signals

The requirements for the ASIC performance are derived from the measured signal current from MIPs in a typical AC-LGAD. Signals for AC-LGAD of 20 and 50 μm thickness, with an internal gain of 20, are shown in Fig.1. It shows the characteristic dependence of LGAD signals on the sensor thickness, i.e. the maximum current does not depend on the thickness, while the rise time and slew rate dV/dt do.

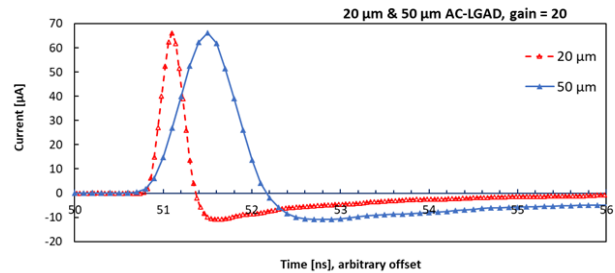

**Figure 1.** Input current from AC-LGAD with gain = 20 measured for 50 μm sensor thickness and scaled for 20 μm sensor thickness.

The signal characteristics critical for the jitter, the time resolution depending on the electronics, are shown in Table 1, which in turn yield the ASIC design goals summarized in Table 2.

Table 1 MIP Signal Characteristics

| LGAD Thickness | 50 μm | 20 μm |
|---|---|---|
| Rise Time (10-90%) [ps] | 455 | 182 |
| Input Charge (G = 20) [fC] | 11 | 4.6 |
| $I_{MPV}$ Input Current [uA] | 15 | 15 |

Table 2 ASIC Design Goals

| ASIC Parameter | 50 μm | 20 μm | Comment |
|---|---|---|---|
| Rise time [ps] | 455 | 182 | Rise time (electronics) = Rise time (sensor signal) |
| Jitter [ps] | 10 | 5 | < 30 % of the "Landau" Noise |
| S/N | > 50 | > 40 | S/N = Rise Time / Jitter |
| Voltage signal [mV] | 70 | 70 | $V_{MPV}=R_{FB}*I_{MPV}$, [$R_{FB}$ = 5 kΩ] |
| Noise RMS [mV] | 1.4 | 1.8 | N = S/(S/N) |
| Internal Sensor Gain | > 20 | > 20 | |

### 2.2 ASIC Design and Layout

The HPSoC design implements signal pre-amplification along with full waveform sampling and digitization in an ultra-small area package size compatible with small-pitch sensors [13]. It is designed to service up to 100 channels in a single die for bidimensional sensor arrays with a pitch as small as 300 μm. The HPSoC will operate with 10 GSa/s waveform digitization and will implement autonomous triggering, feature extraction and multichannel data fusion. The goals for the ASIC design are shown in Table 2. They are based on matching the amplifier rise time to the sensor rise time. The amount of allowable jitter is given by the requirement that the jitter should contribute only a small fraction (<10%) to the total time resolution.

The design and fabrication of a 4-channel HPSoC prototype discussed here emphasizes initially the first stage Trans-Impedance Amplifier (TIA) shown Figure 2. This small prototype ASIC includes a subset of the components which will be required for the full HPSoC design and has been manufactured using a standard 65 nm CMOS process by TSMC. Figure 3 (left) shows a photo of the prototype chip with annotated inputs.

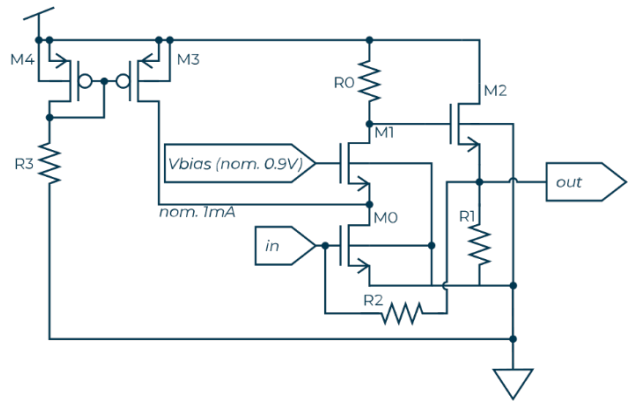

**Figure 2.** Schematic of the HPSoC TIA.

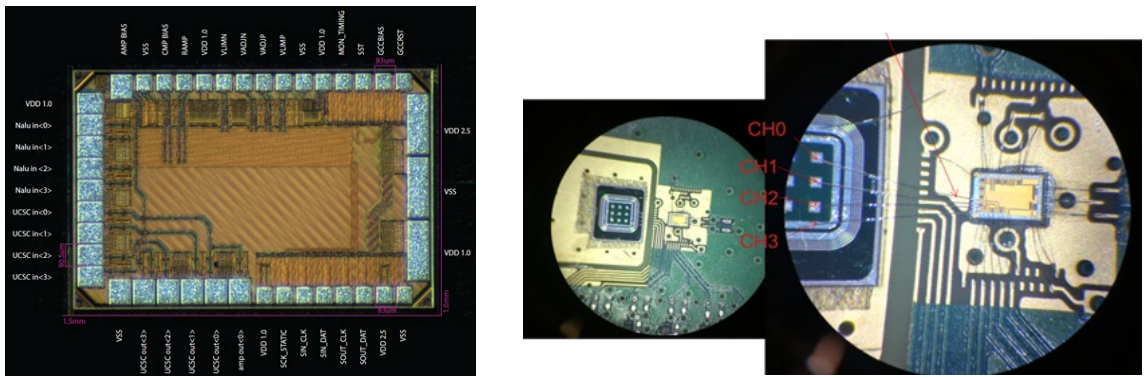

**Figure 3. (left)** Micrograph of fabricated 1.5mmx1.0mm HPSoC prototype chip;
**(right)** Test board with AC-LGAD sensor wire-bonded to HPSoC prototype ASIC.

## 3. Results

The HPSoC ASIC prototype was mounted on a PCB that was custom designed to read out one channel of the chip during electrical calibration, laser TCT and charge collection with β-particles. The TIA was read out with a fast oscilloscope and the signal was analyzed in terms of the collected charge, the maximum of the pulse height (Pmax), the rise time (10-90%) and the RMS noise of the base line. While the calibration was performed without load, for the laser TCT and β-particle charge collection, the ASIC input was wire bonded to a LGAD or AC-LGAD as described below. The arrangement is shown in Fig. 3 (right).

### 3.1 Calibration Results

Then calibration pulse was generated by passing a step pulse of 200 ps rise time and height P through a 0.1 pF capacitor, thus creating an input charge of 0.1* P [fC/V]. The output of the TIA is shown in Fig. 4 (left) when recorded with a differential probe with three different scope bandwidths. While the signals taken at 2.0 GHz and 13.5GHz are indistinguishable, the pulse taken at 500 MHz is clearly slowed down. Thus, the charge collection data was recorded using the 2.0 GHz scope bandwidth, since that setting exhibits lower noise RMS than the 13.5 GHz setting. Figure 4 (center) shows the signal amplitude as a function of the input charge: it is linear up to 100 fC input and its slope is reduced to 1/3 above 200 fC. Figure. 4 (right) shows a rise time as low as 300 ps, the same order of magnitude as the rise time of the input pulse mentioned above.

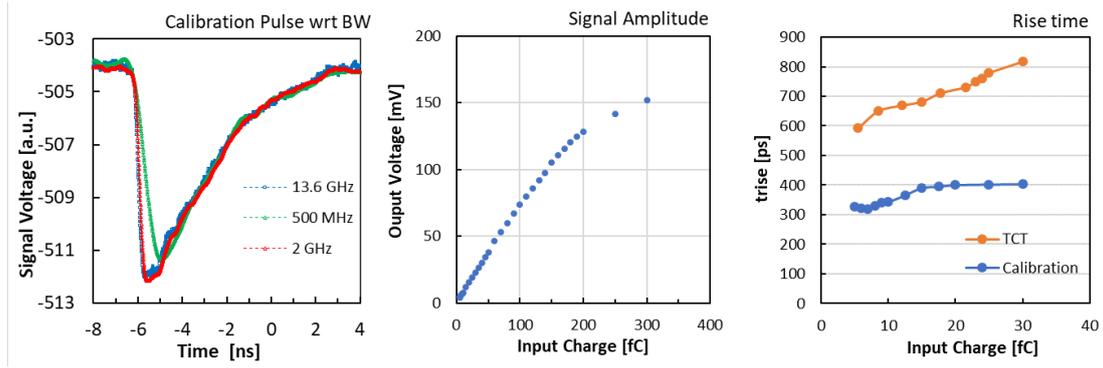

**Figure 4.** TIA output: **(left)** calibration pulse recorded with scope bandwidth 500 MHz, 2.5 GHz and 13.5 GHz; **(center)** calibration Pmax vs. input charge, **(right)** 10-90% rise time, both for calibration and TCT.

### 3.2 Laser TCT Results

For the laser TCT one channel of the ASIC was wire-bonded to a 60 μm thick 3x3 AC-LGAD array from the FBK RSD1 production (500 μm pitch, 200x200 μm$^2$ metal pads, capacitance ~ 120 fF). The IR laser beam was focused close to the connected pad and the input signal was varied by increasing the sensor gain with the bias voltage. As shown in Fig. 4 (right), the observed rise time for TCT of between 600 and 800 ps is about twice rise time from the calibration, since it includes the rise time of the sensor signal.

### 3.3 β-Scope Results

The response to β particles was tested to determine the time jitter, calculated event-by-event from the rise time Trise and the signal-to-noise ratio SNR = Signal/Noise [1]:

$$Jitter = \text{Noise}/(dV/dt) \approx \text{Trise/SNR}$$

This was done on two different LGAD types: the timing and pulse shapes were tested with the AC-LGAD described above. To measure the signal height (= Pmax) would require readout of all neighboring pads sharing the created pulse. Therefore, the SNR was determined by comparing the pulses from a single pad of a 50 μm thick TI-LGAD, read out by either HPSoC or the fast single channel discrete component "UCSC" board based on SiGe transistor technology [4]. In both cases the noise was of the order of RMS ≈ 1 mV.

#### 3.3.1 Response to AC-LGAD

The β data using the AC-LGAD was taken with an elevated self- trigger threshold. The pulse shape of the HPSoC shown in the insert of Fig. 5 (left) indicates the very fast rise time and somewhat slower fall-time of the pulse. The main histogram of Fig. 5 (left) is the 10-90% rise time of the AC-LGAD sensor with a mean of 677 ps. Since this was done with a 60 μm thick sensor, we can scale this number by the sensor thickness to arrive at

$$\text{Rise time (50um)} = 677*(50/60) \text{ ps} = 564 \text{ ps},$$

which is within 20% of the goal in Table 2.
The jitter distribution in Fig.5 (right) indicates the expected 1/Pmax dependence.

#### 3.3.2 Response to TI-LGAD

Although the HPSoC pulse shape in the insert of Fig. 5 (left) appears to be very sharp, a comparison with the TI-LGAD pulses using the UCSC board in Fig. 6 shows the longer tail and the lower signal height of the HPSoC, even though the rise times are comparable. The pulse height ratio is about 4 and the pulse width ratio about 1/3. Using the value of the pulse height Pmax = 28mV from Fig. 6 (left), we can estimate a HPSoC jitter of 45 ps. A redesign of the TIA to "sharpen" the backend of the pulse to gain a factor 3 in pulse height would result in Pmax = 84 mV and a jitter of between 10 to 20 ps.

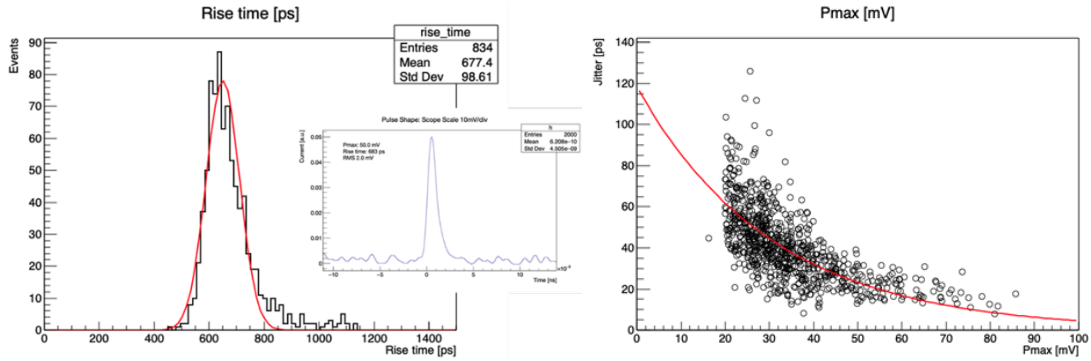

**Figure 5.** HPS0C β-pulse data from a 60 µm thick AC-LGAD: **(left)** Rise time with a mean of of 677 ps (insert: pulse shape); **(right)** observed jitter vs Pmax.

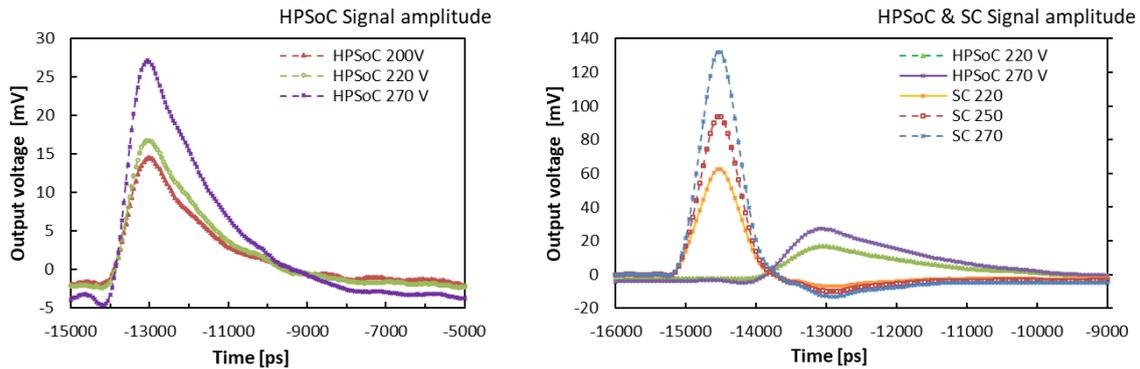

**Figure 6.** Average β-pulse distributions from a 50 µm thick TI-LGAD sensor read out by: **(left)** HPSoC prototype, and **(right)** single-channel UCSC board ("SC") compared to HPSoC prototype
(with arbitrary delay, note the different time and voltage scales).

## 4. Conclusions

The HPSoC prototype shows very good rise time: 300 ps calibration, 564 ps for 50 µm sensor.
Lower gain and wider pulse width are observed with respect to circuit simulation, which is currently being investigated.
Future work will include the optimization of the input stage, such as lowering the input impedance, co-optimization for the digitizer load target, and optimization for different sensor geometries, i.e. longer strips.

**Acknowledgments**

This work was partially funded by US DOE SBIR Grant DE-SC0021755 (Nalu) and US DoE Grant DE-SC0010107-005 (SCIPP).